\documentstyle[12pt,aas2pp4]{article}
\begin{document}

\title{THE CHEMICAL EVOLUTION OF STARBURST NUCLEUS GALAXIES}

\author{R. Coziol\altaffilmark{1,4}, T. Contini\altaffilmark{2}, E. Davoust\altaffilmark{2}}
\and 
\author{S. Consid\`ere\altaffilmark{3}}

\altaffiltext{1}{Laborat\'orio Nacional de Astrof\'{\i}sica - LNA/CNPq, Rua Estados Unidos, 154, Bairros das Nac\~oes - Itajub\'a/MG (Brasil), CEP 37500-000}
\altaffiltext{2}{UMR 5572, Observatoire Midi--Pyr\'en\'ees, 14 avenue E. Belin, F-31400 Toulouse, France}
\altaffiltext{3}{EP CNRS 123, Observatoire de Besan\c{c}on, B. P. 1615, F-25010 Besan\c{c}on cedex France}
\altaffiltext{4}{On Leave of absence from the Instit\'uto Nacional de Pesquisas Espacias/INPE}

\begin{abstract}
The metallicities derived from spectroscopic observations of a sample of 
Starburst Nucleus Galaxies (SBNGs) are compared to those of several other 
types of galaxies (normal giant galaxies, Irregular and HII galaxies) drawn 
from the literature.  The SBNGs are deficient in metals with respect to normal 
galaxies of same morphological type, suggesting that {\it SBNGs are galaxies 
still in the process of formation}.  

Breaking the SBNGs into early--types (Sb and earlier) and late--types 
reveals that the former seem to follow the same linear luminosity--metallicity 
relation as the irregular and elliptical galaxies, whereas the latter and the 
giant spirals show comparable (0.2 and 0.3 dex) excess abundances with 
respect to the linear relation.  This difference between the two types 
of SBNGs is consistent with the predictions of the model of hierarchical 
formation of galaxies: the early-type SBNGs are building their bulges by 
successive mergers of small stellar and gaseous systems, while the late--type 
SBNGs are mostly accreting gas to form a disk.  
 
\end{abstract}


\section{Introduction}

The elemental abundance in galaxies is an essential parameter for 
understanding how galaxies evolve, and provides important constraints on 
current models for their formation. It also constrains 
models for stellar evolution and for primordial nucleosynthesis.  

The elemental abundance in spiral galaxies was first thought to depend 
essentially on local properties of the interstellar medium, such as the
density of gas.  Recent studies have shown that it 
depends predominantly on global properties of the galaxies, such as mass or 
Hubble type, but the situation is still rather confused.  Garnett \& Shields 
(1987) find evidence for a correlation between mean metal abundance and total 
luminosity. Oey \& Kennicutt (1993) find no such correlation for 
early-type (Sa to Sb) galaxies, but note that early-type
galaxies have higher metal abundances than late-type ones.
Using a different sample of galaxies, Zaritsky, Kennicutt \& Huchra (1994; 
hereafter ZKH) confirm that the abundance is correlated with total 
luminosity; they also state that stochastic events, such as starbursts and 
external accretion of matter, could only contribute to the scatter in the 
abundance values. 

A first evidence that starbursts are in fact a driving mechanism in the 
chemical evolution of spiral galaxies has been presented by Coziol (1996b).
He finds that the Starburst Nucleus Galaxies (hereafter SBNGs) are less 
chemically evolved than galaxies with similar morphologies and comparable 
luminosities, and that the SBNGs seem to follow the same 
luminosity--metallicity linear relation as the irregular and elliptical 
galaxies. 

In this {\sl Letter}, we use new spectroscopic data on a large sample 
of Markarian SBNGs (Contini 1996) to establish Coziol's claim on firmer 
observational ground. In other words, we compare the global chemical 
characteristics of early- 
and late-type SBNGs with those of normal giant 
galaxies, of HII galaxies, of irregular and elliptical galaxies.
 
\section{The different samples}

Our sample of SBNGs is composed of 62 Markarian barred galaxies from the study 
of Contini (1996), of a sample of 40 compact Kiso galaxies (Comte et al. 1994) 
and of a sample of 20 SBNGs from the MBG survey (Coziol et al. 1994, 1996). 
The starburst nature of all these galaxies was established by the different 
authors in the original articles using standard diagnostic diagrams of emission 
line ratios (Baldwin et al. 1981, hereafter BPT; Veilleux \& Osterbrock 1987, 
hereafter VO). 

The comparison samples were taken from the literature. We used the sample of 
normal spiral galaxies of ZKH, to which we added the sample of early--type 
spirals observed by Oey \& Kennicutt (1993).  The sample of irregular galaxies 
is from Skillman et al. (1989). As samples of 
HII galaxies (see Coziol 1996a, for a definition of the two main types of 
starburst galaxies), we used those of the Cal\'an-Tololo survey (Pe\~na et al. 
1991) and those of the catalogue of Terlevich et al. (1991). Finally, we also 
included a sample of luminous Arp interacting galaxies (Keel et al. 1985). 

Since we are using various sources for our samples of galaxies, we must verify 
the consistency of the derived abundances. Those of the giant spiral galaxies 
are average values; they were estimated by measuring the mean values, 
normalized to a mean radius, of the abundances of at least 10 disk HII 
regions. For the starburst galaxies, the measures were done with long slits 
centered on the nucleus (SBNGs) or on the most luminous part of the galaxy 
(HII galaxies). Because the HII galaxies have small angular dimensions, the 
slit aperture usually covers the entire galaxy and the measured metallicities 
are therefore good estimates of their mean metallicities.  In the case of the 
SBNGs, which have higher angular dimensions than the HII galaxies, the 
abundances are mostly those of the nuclei. In normal spirals, the 
metallicities usually increase toward the center of the galaxies (ZKH); 
preliminary results indicate that this is probably also true for the SBNGs 
(Consid\`ere et al. in preparation). The abundances of the SBNGs represent 
therefore upper limits of their mean metallicities.  

The metallicities ([O/H]) of the HII galaxies were estimated  by determining 
the electron temperature using [OIII]$\lambda$4363. For the SBNGs this line is 
usually not observed and we used the metallicity index R$_{23}$ (Pagel et al. 
1979), or some comparable methods based on R$_{23}$. We verified that all 
these methods give similar results. One of the methods (Coziol et al. 1994) is 
based on the calibration of a diagnostic diagram using HII region samples 
where the electron temperature is determined using 
[OIII]$\lambda$4363 and covering metallicities between -0.9 and 0.3 dex. The 
differences between the metallicities obtained by this method and the others 
are generally much lower than the typical uncertainty of 0.2 dex associated to 
each of these methods. 

The redshifts and the adopted morphologies for all the galaxies are as given 
in the original papers or were found in NED\footnote{The NASA/IPAC 
Extragalactic Database} or in LEDA\footnote{Lyon Meudon Extragalactic 
database}. 
For some galaxies, the B magnitudes were not given in the original 
papers and were found in NED. 
No internal extinction correction was applied. 
All the absolute magnitudes were determined or corrected for the value H$_o = 
75$ km s$^{-1}$ Mpc$^{-1}$. 

\section{Results}

Table 1 gives the mean absolute magnitudes and metallicities for the different 
samples of galaxies. The early--type SBNGs are poorer in metals on average by 
0.2 dex with respect to the late--type SBNGs and by nearly 0.3 dex with 
respect to the giant spirals. On the other hand, both types of SBNGs share the 
same distribution in luminosities. This is a strong indication that the 
differences between the two types of SBNGs are real, and not a selection 
effect. Moreover, the low abundance cannot be an artefact caused by the small 
aperture used in our spectroscopic setup, since this tends to raise, rather 
than lower, the measured abundance. 

In figure 1, we compare the behavior of the abundance with respect to 
morphology for the SBNGs and the normal and HII galaxies. 
As expected, the normal galaxies show a morphology--metallicity trend: the 
metallicity increases from the late--type spirals to the early--type ones. 
This trend is not observed for the HII galaxies, which have lower 
mean abundances than normal galaxies; this is not new and is 
related to the low luminosities and small 
masses of these galaxies. 
But our luminous SBNGs do not follow the morphology--metallicity 
trend either. Although chemically richer than the HII galaxies, the SBNGs 
are nonetheless significantly deficient in metals as compared to normal 
galaxies. In figure 1, this phenomenon is most evident for the early-type 
SBNGs (T $\leq 3$).

Our results are based on the assumption that the main source of ionization of 
the gas in the SBNGs is OB stars, like in normal HII regions. This is implicit 
in the definition of starbursts based on  diagnostic diagrams of emission line 
ratios. But the spectra of the SBNGs show one important difference 
with respect to those of disk HII regions: their ratios [NII]/H$\alpha$ are on average 
0.2 dex higher (Coziol et al. 1996). If this excess emission corresponds to a 
supplementary nonthermal ionizing source, like a hidden AGN or shock--heated 
gas, such as embedded supernova remnants, this could indeed produce a false 
effect of lower metallicity. 

To verify this hypothesis, we compare in figure 2 the ratios [SII]/H$\alpha$ 
with the ratios [NII]/H$\alpha$. Both ratios would increase in the presence of 
a nonthermal ionizing source. We find no relation between the ratios of these 
two lines. We also verified that there is no relation between [SII]/H$\alpha$ 
or [NII]/H$\alpha$ and the metallicity. 
In figure 2, the dot--dashed line corresponds to the lower  
limit predicted by shock models (see VO and Ho et al. 1993). The values for 
the SBNGs are well below this value. In fact, the majority of SBNGs have 
[SII]/H$\alpha$ ratios within 0.2 dex of the mean ratios observed in normal 
HII regions (Greenawalt \& Walterbos  1996). In figure 2, we 
also see that very few SBNGs have Nitrogen emission above the lower 
limit proposed by Ho et al. (1993) for transition galaxies (that is galaxies 
with AGN $+$ HII region spectra). 
V\'eron et al. (1996) also found very few transition galaxies. 
The presence of a hidden AGN in the SBNGs is also ruled out because of the 
weakness of [OI]$\lambda$6300: only $\sim$ 40\% of our galaxies show this line 
and their ratios Log([OI]/H$\alpha$) $\leq -1.3$ are similar to those of 
normal HII regions (BPT, VO). 
We conclude that there is nothing in the spectra of the SBNGs to 
prevent us from applying normal HII region models to SBNGs, and the low 
metallicity of the SBNGs is therefore real. 

A first assumption that comes to mind for explaining the low metal abundance 
in SBNGs is massive accretion of unprocessed gas during gravitational 
interaction with another galaxy (Coziol 1996b), which would explain both the 
reduced elemental abundance and the starburst. But most SBNGs are isolated 
(Contini 1996; Coziol et al. 1996) and there is no evidence of lower 
abundances in the sample of luminous Arp interacting galaxies.
This suggests that the low--metallicity is a characteristic of the SBNGs, 
which depends on their particular history of formation. 

In figure 3, we show the diagram of metallicity as a function of luminosity 
for the galaxies of our samples. The solid line is the linear relation for the 
irregular and elliptical galaxies reported by ZKH. It appears that the 
early--type SBNGs scatter around this relation more closely than do the other galaxy types. To test this hypothesis, 
we calculated the differences between the abundances predicted by the linear 
relation and the observed abundances for each of the galaxies. The 
distributions of these differences are shown in figure 4. The early-type SBNGs 
have the same distribution as the irregulars, whereas the late--type SBNGs 
show the same type of deviation as the giant spirals. The hypothesis that the 
distributions for the early and late--type SBNGs come from the same parent 
population is rejected at a confidence level of 99\% with a 
Kolmogorov--Smirnov test. 
The deviation of the HII galaxies from the linear relation in figure 3 and 4 
is a consequence of the starbursts, because their luminosity is more affected 
by starbursts than that of massive galaxies,  
and they are therefore more luminous than their metallicity suggests.  

\section{Discussion}

Our new data on SBNGs allow us to confirm the phenomenon discovered by Coziol 
(1996b): the SBNGs are less chemically evolved than normal galaxies. This is 
inconsistent with the current hypothesis that SBNGs are evolved galaxies which 
were rejuvenated by interactions.
 
The difference between the early-- and late--type SBNGs is also very 
meaningful, because, among the mechanisms considered by Coziol (1996b) for 
explaining the metal deficiency of SBNGs, only one predicts such a difference. 
This is the model of hierarchical formation of galaxies (Tinsley \& Larson 
1979), according to which ellipticals and bulges of spiral galaxies 
are formed by a sequence of mergers of stellar and gaseous systems. Therefore, 
bulge-dominated SBNGs must follow the same luminosity--metalli\-city 
relation as ellipticals. This implies that bulges of spirals are 
similar to elliptical galaxies, which is now supported by observations (see 
Jablonka et al. 1996). 
 
If a sufficiently large fraction of gas is left from the initial 
merger, it will collapse to form a disk. Struck--Marcell (1981) showed that
when the gas fraction of matter accreted increases,   
the successive generations of star   
have higher abundances than in the merger case.  
During galaxy formation, the metallicity of the gas will therefore increase 
faster in the disk--dominated SBNGs than in the bulge--forming SBNGs. It is 
very interesting to find that the disk--dominated SBNGs share the same 
position in the luminosity--metallicity diagram as the giant spirals. Indeed, 
following Kennicutt (1983), one explanation for the nearly constant star 
formation rates of the giant spiral galaxies over the last few Gyrs is that 
they have accreted extra gas in their disk.  

\section{Conclusion}

We confirm Coziol's (1996b) claim that luminous SBNGs are less chemically 
evolved than normal spiral galaxies. This is a strong indication that 
luminous SBNGs are in fact galaxies still in the process of formation. 
We have also found a difference between the abundances of the early-- and 
late--type SBNGs. This difference is consistent with the predictions of the 
model of hierarchical formation of galaxies. Our results suggest that galaxy 
formation is a continuing process and that the starburst phenomenon is a 
normal phase in the formation of all galaxies.

This scenario has some important implications for the observation of galaxies 
in formation at high redshifts. It predicts that the fraction of interacting 
and merging objects should be higher in the past; at high redshifts the number 
of early--type galaxies should be lower and the grand--design late--type 
spirals should appear later. These predictions may already have been verified 
by observations (Kauffman et al. 1996; van den Bergh et al. 1996).   

\acknowledgements
R. Coziol acknowledges the financial support of the FAPESP under contract 94/3005-0 and of the CNPq, under contracts 360715/96-6 (NV). 
This research has made use of the 
NASA/IPAC Extragalactic Database (NED) which is operated by the Jet Propulsion 
Laboratory, California Institute of Technology, and of the Lyon Meudon 
Extragalactic database (LEDA), supplied by the LEDA team at CRAL-Observatoire 
de Lyon (France).  We thank Sylvain Blaize for his contribution to this 
project.

\clearpage
\begin{deluxetable}{lccc}
\footnotesize
\tablecaption{Absolute magnitudes and metallicities of 
the samples\label{tbl-1}}
\tablewidth{0pt}
\tablehead{
\colhead{sample}&\colhead{N}&\colhead{M$_B$}&\colhead{[O/H]}
}
\startdata 
Giant spirals&53&--20.28 $\pm$  1.43&\phs 0.03  $\pm$  0.05\nl 
SBNG (early) &76&--19.68 $\pm$  1.28&--0.25  $\pm$  0.04\nl
SBNG (late)  &46&--19.74 $\pm$  1.66&--0.05  $\pm$  0.08\nl 
Irregulars   &20&--15.20 $\pm$  5.09&--1.06  $\pm$  0.13\nl 
HII galaxies &44&--17.64 $\pm$  2.93&--0.86  $\pm$  0.07\nl
\enddata 
\end{deluxetable}

\clearpage

\figcaption[cozetal.fig1.ps]{Mean metallicity as a function of morphology for 
SBNGs, normal galaxies and HII galaxies. The luminous SBNGs are 
metal deficient compared to normal galaxies, but this is not the case for the 
Arp interacting galaxies.} 

\figcaption[cozetal.fig2.ps]{Diagram of [SII]$\lambda\lambda$6717,6731 as a 
function of [NII]$\lambda$6584. Dash--dotted line $=$ lower limit for shock 
models; dashed line $=$ mean of normal HII galaxies; vertical dotted line $=$ 
limit for transition type galaxies. There is no indication of an additional 
source of ionization.} 

\figcaption[cozetal.fig3.ps]{Luminosity--metallicity diagram for the different 
types of galaxies.  Bender \& Huchra's linear relation is for 
elliptical galaxies and is as given in ZKH (1995). It is mostly the 
early--type SBNGs that follow the linear relation for elliptical and 
irregular galaxies.} 

\figcaption[cozetal.fig4.ps]{Dispersion of the deviations from the linear law 
([O/H]$_{th} -$[O/H]$_{obs}$). The dispersion for the early--type SBNGs is the 
same as for the irregulars. The late--type SBNGs exceed the predicted 
abundances in nearly the same way as the giant spirals.} 

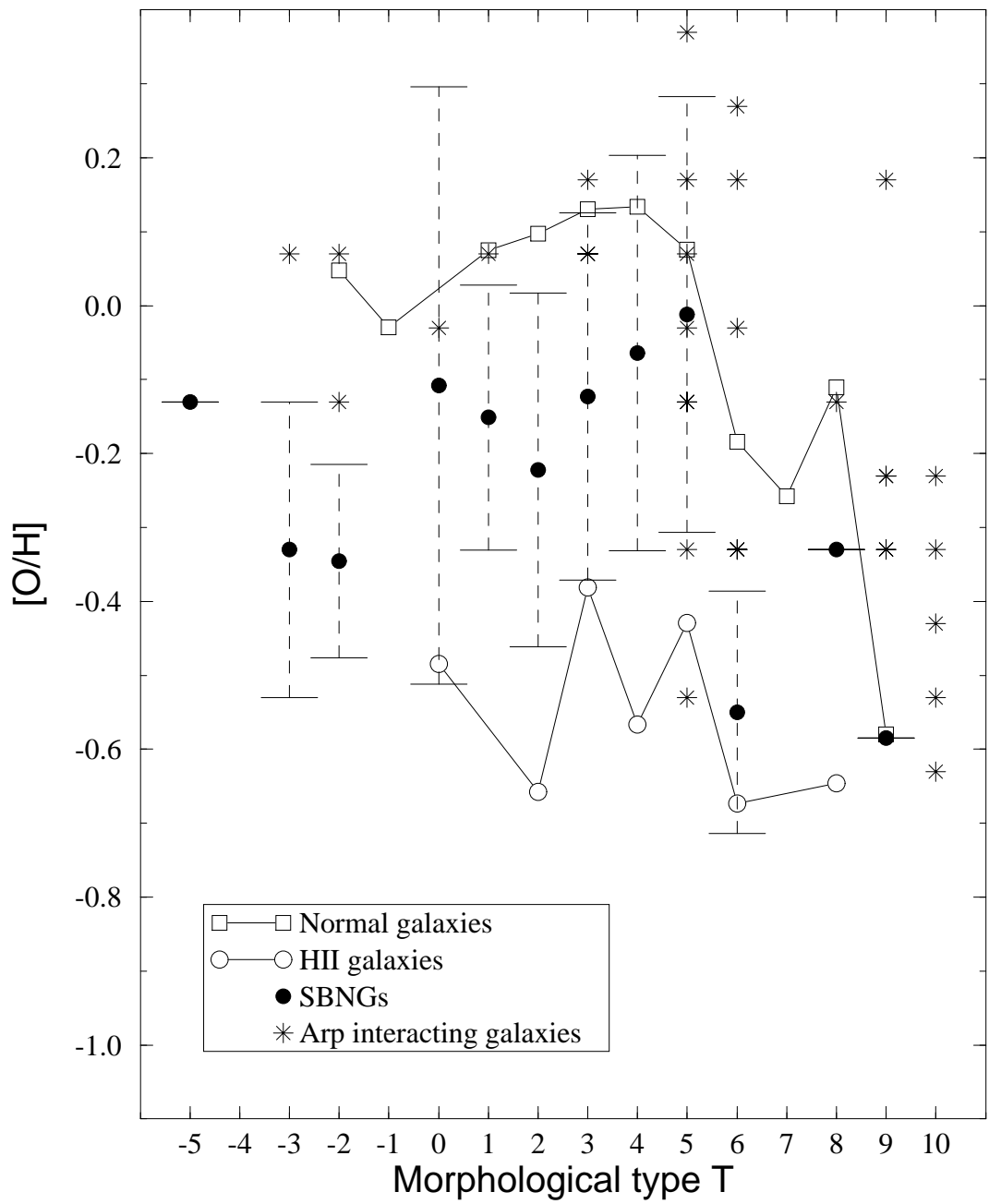
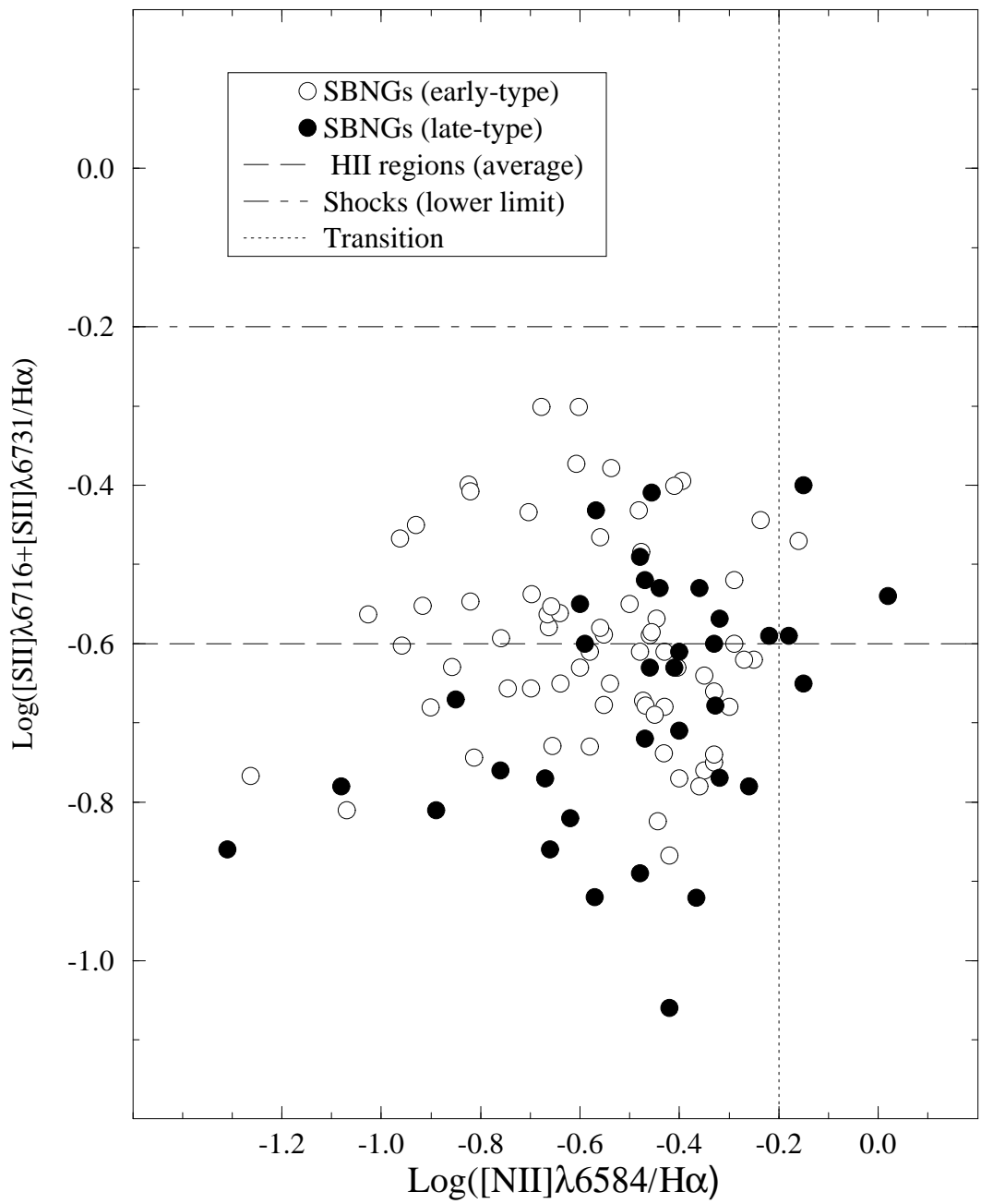
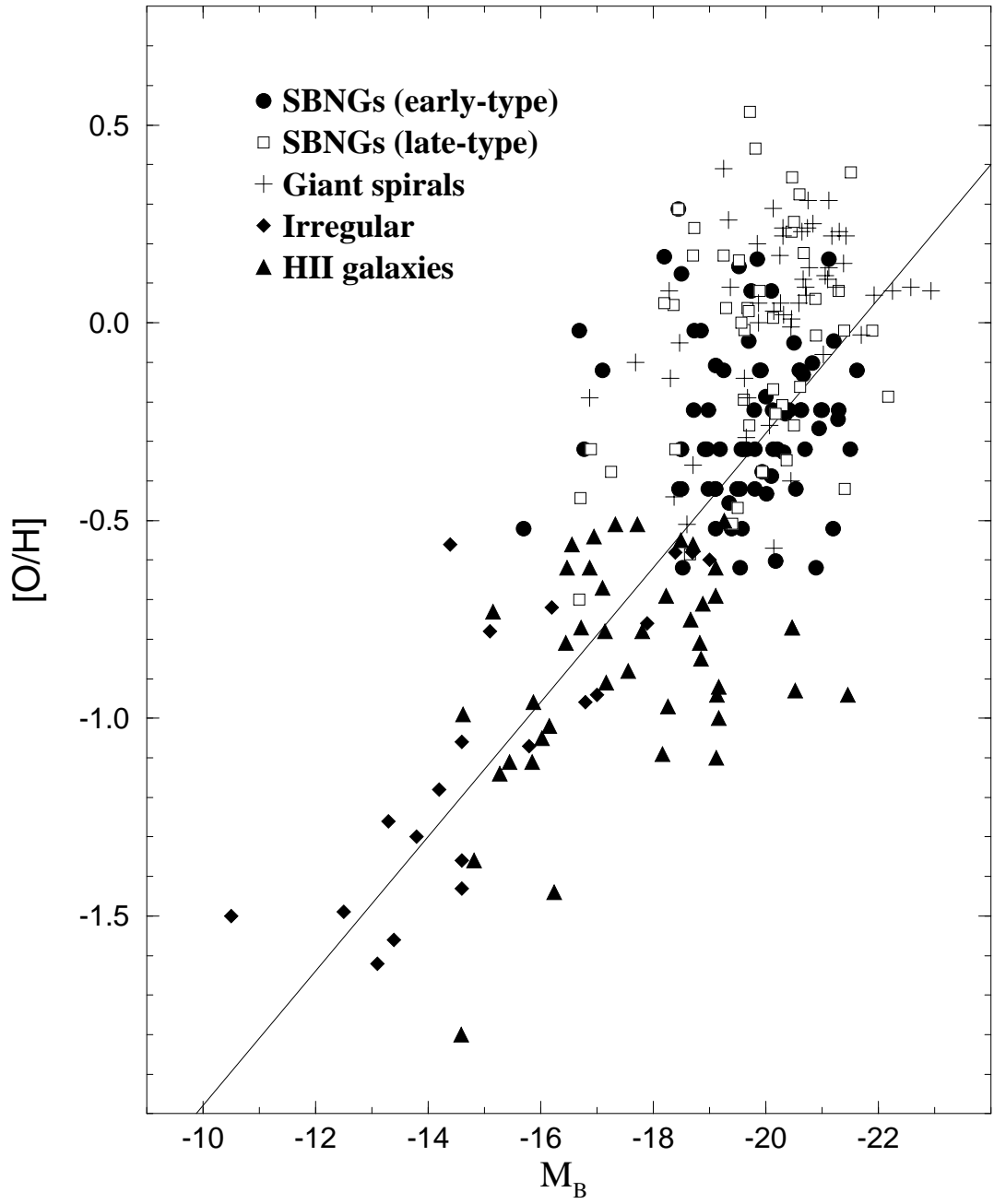
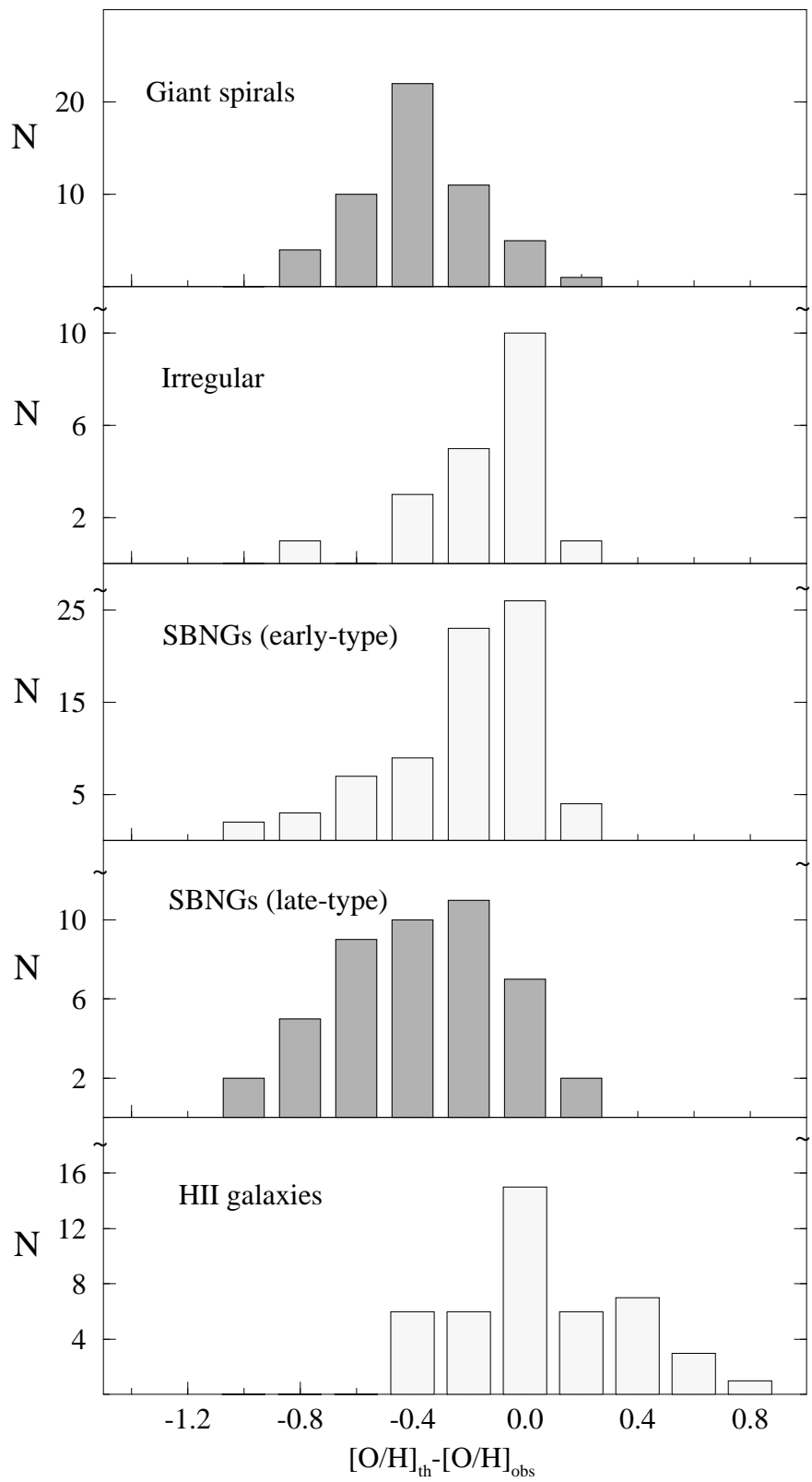
\end{document}